# Performance of the combined zero degree calorimeter for CMS.

**O A Grachov**[1]**, M Murray, J Snyder, J Wood, V Zhukova**
Departments of Physics and Astronomy, University of Kansas, Lawrence, KS, USA

**A S Ayan, P Debbins, D F Ingram, E Norbeck, Y Onel**
Department of Physics and Astronomy, University of Iowa, Iowa City, IA, USA

**E Garcia**
Department of Physics and Astronomy, University of Illinois at Chicago, Chicago, IL, USA

**G Stephans**
Laboratory for Nuclear Science, Massachusetts Institute of Technology, Cambridge, MA, USA

**For CMS Collaboration**

E-mail: grachov@ku.edu

**Abstract.** The combined zero degree calorimeter (ZDC) is a combination of sampling quartz/tungsten electromagnetic and hadronic calorimeters. Two identical combined calorimeters are located in the LHC tunnel at CERN at the straight section ~140 m on each side of the CMS interaction vertex and between the two beam pipes. They will detect very forward $|\eta| \geq 8.5$ photons and neutrons. ZDC information can be used for a variety of physics measurements as well as improving the collision centrality determination in heavy-ion collisions. Results are presented for ZDC performance studies with the CERN SPS H2 test beam.

## 1. Introduction
Two identical zero degree calorimeters (ZDCs) [1], will measure very forward neutrons and photons in heavy-ion and early p-p collisions and will cover pseudo rapidity region from 8.5 (a horizontal angular width of 580 μrad). The ZDC are located 140 m on each side of the CMS interaction vertex (IP5). They are between the two beam pipes inside the neutral particle absorber (TAN) downstream of the first beam dipole magnet at the straight section.
The number of spectator neutrons can be used for beam tuning and luminosity monitoring, and for determination of the impact parameter or centrality of interaction in heavy-ion collisions. With an additional shower maximum detector [2], the ZDC will measure of the reaction plane of A-A and p-A collisions. The detector is useful for study of diffractive processes in p-p collisions [3] and also, as a neutron tagging detector for selection of γ-γ and γ-A collisions in ultra-peripheral heavy ion collisions [4]. The location of one ZDC calorimeter in LHC sector 5-6 is presented on figure 1.

---
[1] To whom any correspondence should be addressed.

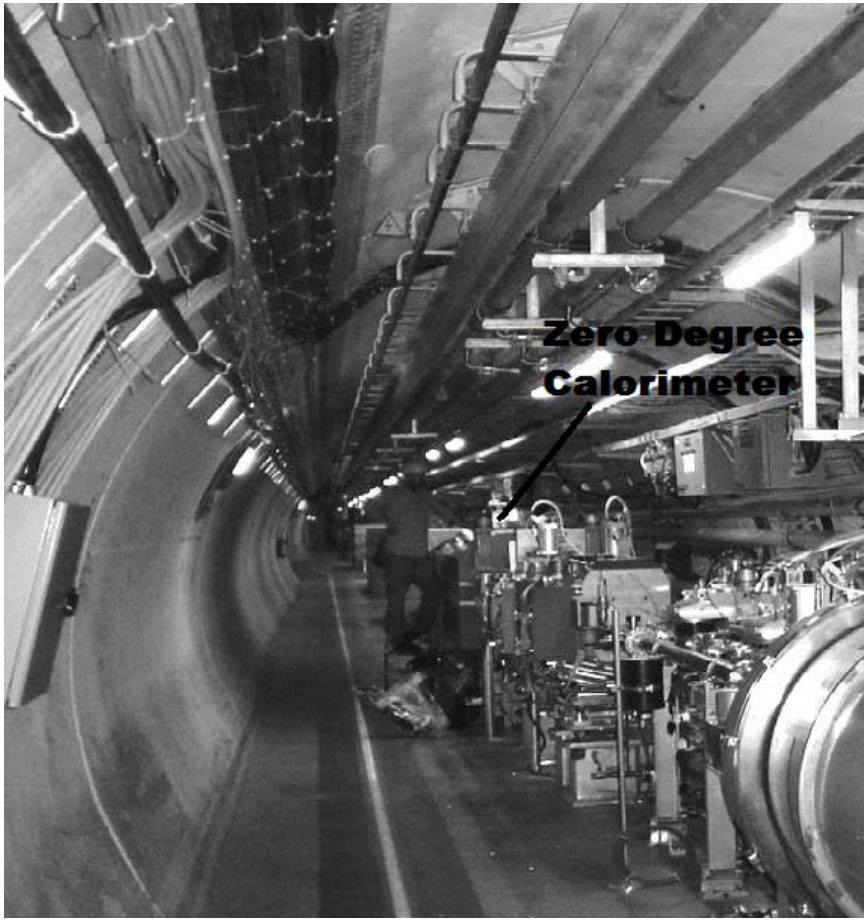

**Figure 1.** Photo presents a part of LHC sector 5-6 at the location of one ZDC detector. The detector is located in a narrow space inside the neutral particle absorber (TAN) at 140 m from IP5.

## 2. Requirements
The calorimeter design was dictated by physics goals and by conditions at their location in the LHC tunnel. Specifically:

- The calorimeter should allow reconstruction of the energy of 2.75 TeV spectator neutrons and 50 GeV photons with a resolution of 10-15%;
- Radiation hardness consideration.
During the early p-p runs and design luminosity Pb-Pb runs, the expected average absorbed radiation doses will be approximately 180 MGy and 300 kGy, respectively, per data-taking year;
- The calorimeter should utilize fast detection technology.
Proton bunch crossing every 25 ns (40 MHz)..
The expected interaction rate for minimum bias event in Pb-Pb collisions is ~8000 event/sec ;
- Very compact device.
The calorimeter must fit in a space with dimensions: 96mm (W) x 607mm (H) x ~700mm (L)

## 3. Calorimeter design

A compact sampling quartz fiber calorimeter with W (tungsten)-radiators [1] meets all requirements. The design of each individual ZDC includes two independent sampling calorimeter sections: an electromagnetic section (EM section) and a hadronic section (HAD section). Each section consists of a tungsten plate/ quartz fiber ribbon stack.

The HAD section is longitudinally segmented into 4 readout segments (towers) of ~1.4 nuclear interaction length each. The electromagnetic section (19 radiation length or ~ 1 nuclear interaction length) segmented into 5 horizontal individual readout towers with transverse size of 15.56 mm. The ZDC EM and HAD section use a 90° or 45° – orientation geometry respectively (see figure 2 and figure 3).

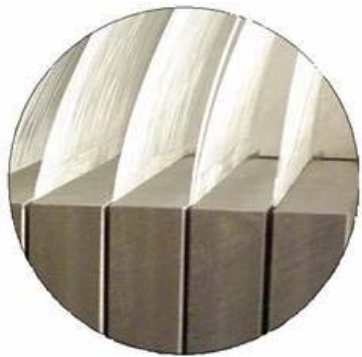 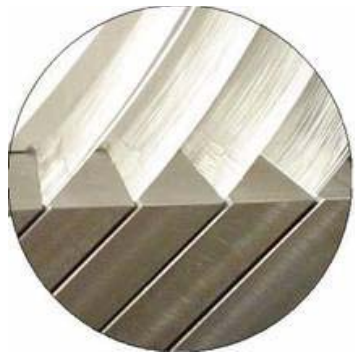

**Figure 2.** 90° - orientation geometry for electromagnetic section .

**Figure 3.** 45° - orientation geometry for hadronic section.

Both sections have transverse dimensions of 79.3 mm and height of 100 mm. This small aperture is sufficient to capture all fragmentation neutrons in peripheral collisions. The layout and numbering scheme of ZDC detectors is presented on figure 4. The coordinate definitions are the same as the CMS coordinate system.

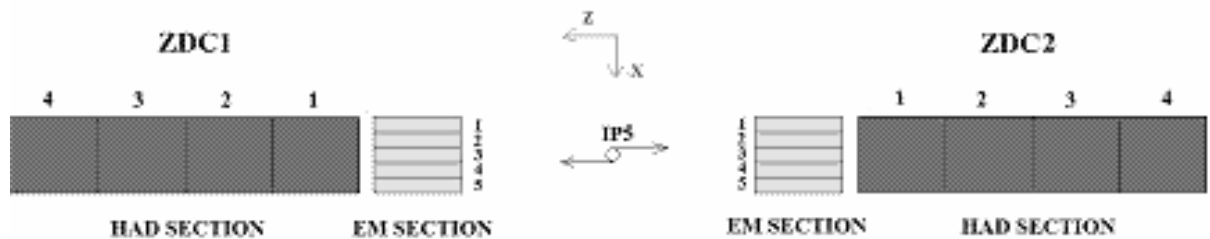

**Figure 4.** Layout and numbering scheme of ZDC detectors.

## 4. Experimental setup

The main goals of the test beam measurements were to study the energy resolution and linearity of the zero degree calorimeter. It also tested the electronic chain to be used with the CMS. Measurements were carried out in the SPS H2 beam at CERN in August 2006 and May 2007.

A 400 GeV/c proton beam was extracted from the SPS accelerator and steered onto the primary target. Downstream from the primary target, a secondary beam was made with momenta from 10 to 350 GeV/c. The resulting beam then passed through a gas threshold Cherenkov counter, multi-wire proportional chambers and scintillating counters (see figure 5).

The trigger system required a coincidence between the scintillating counters and the appropriate signal in the Cherenkov counter. The normal beam spot exposed an area on the calorimeter of $(2 \times 2)$ cm$^2$.

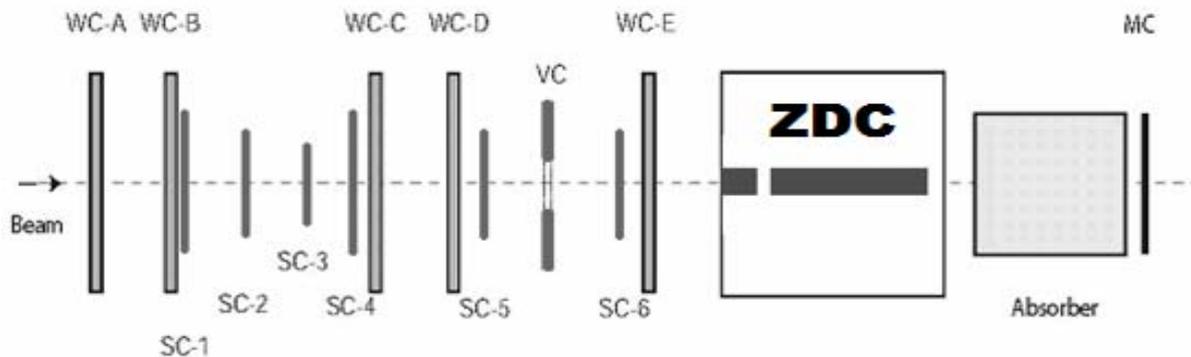

**Figure 5.** Test beam experimental setup.

The calorimeter was mounted on a table that could be moved in horizontal and vertical directions under remote control. This made it possible to position the beam anywhere on the front face of the calorimeter.

## 5. Read-out electronics

There are a total of 18 readout channels for the two ZDCs. Photomultiplier tubes (PMT) are used as photo detectors. The PMT is an 8 - stage Hamamatsu R7525 phototube with bi - alkali photocathode, with average quantum efficiency for Cherenkov light of approximately 10%.

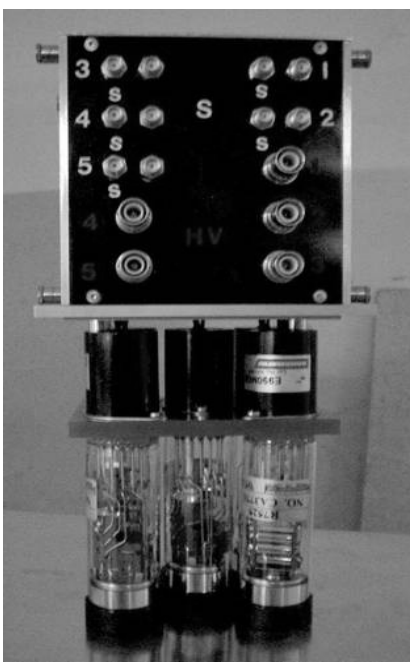

The resistive chain high voltage base (ratio "B") is optimized to achieve a high gain and the maximum dynamic range of linearity.

For future work, the high-voltage for the PMTs will be provided by commercial CAEN NIM units. Five power supplies are mounted in a rack in NIM crate that resides in the counting room. These units will be controlled and monitored via HCAL slow controls.

The PMT's HV base has two connectors: a coaxial signal output connector and a SHV high voltage connector.

The PMT and base are housed in a shielding enclosure, so called PMT read-out box. A photograph of the EM section PMT read-out box is shown on figure 6.

**Figure 6.** PMT read-out box of the EM section

For data collection with CMS, signals from the PMT will be transmitted through a long (204 m, type C–50–11–1) coaxial cable to the patch panel mounted on the 9U VME crate located in the underground counting room (Electronic Rack S2E13). The electronic scheme of PMT signal transmission to patch panel and QIE is presented on figure 7. The patch panel splits each input signal into two output signals: S1, S2 with ratio S1 / S2 = 10/1. S1 provides the coaxial signal output connectors for the distribution of these signals to front - end modules contained QIE (Charge Integrator and Encoder) ADC chips for digitization. S2 is for monitoring and tuning of the interaction of beams and defining the real - time luminosity.

Trigger and front-end electronics run at 40 MHz, the maximum clock speed that will be used at the LHC.

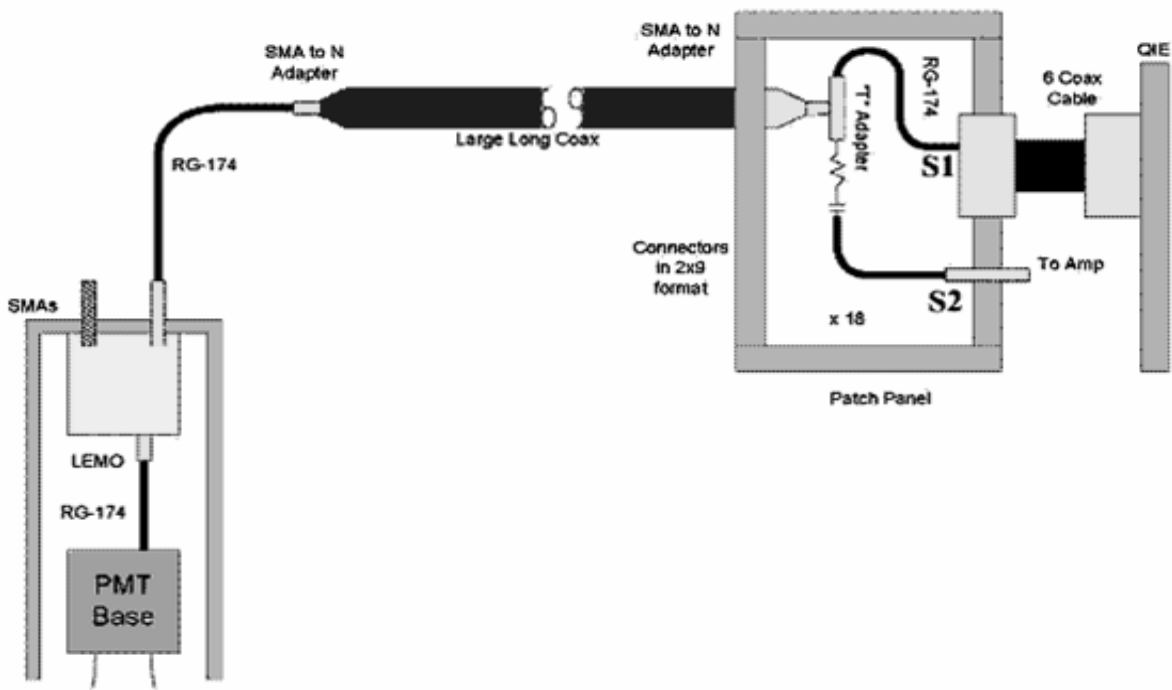

**Figure 7.** Electronic scheme of PMT signal transmission to patch panel and QIE

## 6. Response to positrons

The characteristics of the EM section were measured with positrons of seven different test beam energies (from 10 GeV to 150 GeV). To equalize response among towers the geometrical center of each tower (T1, T2, T3, T4 and T5) was irradiated by 50 GeV positron beams. The peak position obtained from a Gaussian fit of the amplitude distribution for each tower was used to determine the calibration coefficients. Energy deposition in the center tower T3 versus energy deposition in the neighboring towers T2 and T4 (figure 8 and figure 9) show approximately the same response, which shows that the towers were calibrated with a precision better than 10%.

The energy resolution for the different positron energies was obtained by Gaussian fits and was parameterized as $(\sigma/E)^2 = 0.49/E + 0.0064$. The GEANT4 simulations [3] reproduced well the EM section test beam data (figure 10). The linearity of the EM section response was defined as the ratio of the fitted mean energy to the beam energy. The calorimeter was found to be linear over a range from 10 GeV to 150 GeV to within 2%-3% (figure 11).

The analysis did not included corrections for dead material, leakage and the non-compensation nature of the calorimeter.

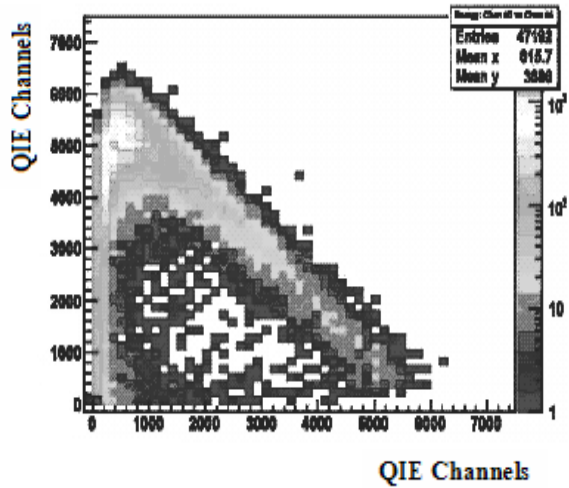

**Figure 8** Energy deposition in tower T2 versus energy deposition in tower T3

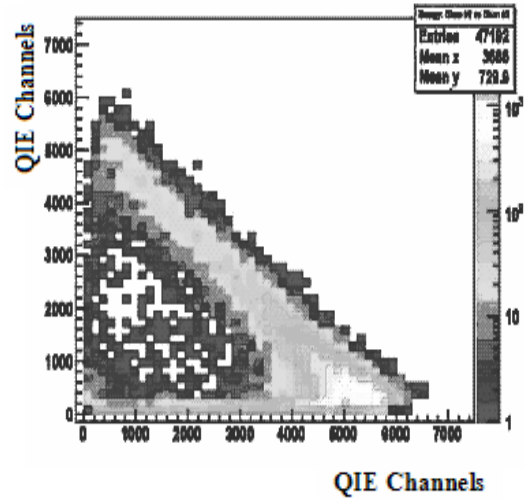

**Figure 9** Energy deposition in tower T3 versus energy deposition in tower T4

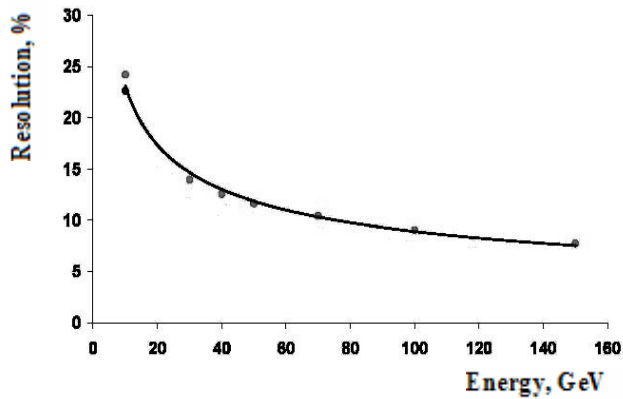

**Figure 10.** Positron energy resolution of EM1 and EM2 electromagnetic sections.

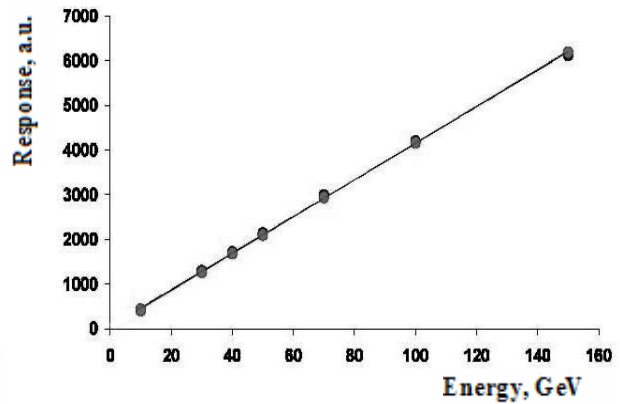

**Figure 11** The linearity response to positrons of EM1 and EM2 electromagnetic sections.

### 7. Response to hadrons

Positive pions with energies of 150, 200, 300 and 350 GeV were used to measure the response of the combined zero degree calorimeter (EM + HAD system). The total depth of the combined system is ~7 hadronic interaction lengths ($\lambda$). The total energy was defined as the sum of the energy depositions in the EM and HAD sections. The energy dependent intercalibration parameter between the EM and HAD section was determined by minimizing the energy resolution for 350 GeV pions. Figure 12 presents energy deposition in the HAD1 section versus of energy deposition in the EM1 section for 350 GeV pions. The energy resolution was obtained by a Landau fit to be 21.5 % for 300 GeV pions and was parameterized as σ/E = 138%/√E + 13% (see figure 13). An extrapolation to energy 2.75 TeV predicts a resolution of approximately 15% (GEANT4 - about 12% [5]). Measurements show good linearity of detector response to hadrons in the range of 100 to 350 GeV.

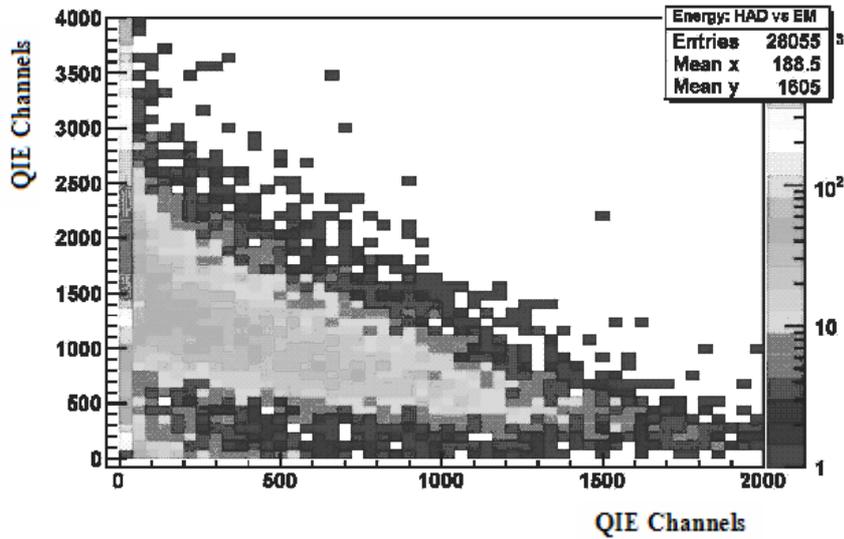

**Figure 12.** Energy deposition in HAD section versus energy deposition in EM section of ZDC1 for 350 GeV positive pions.

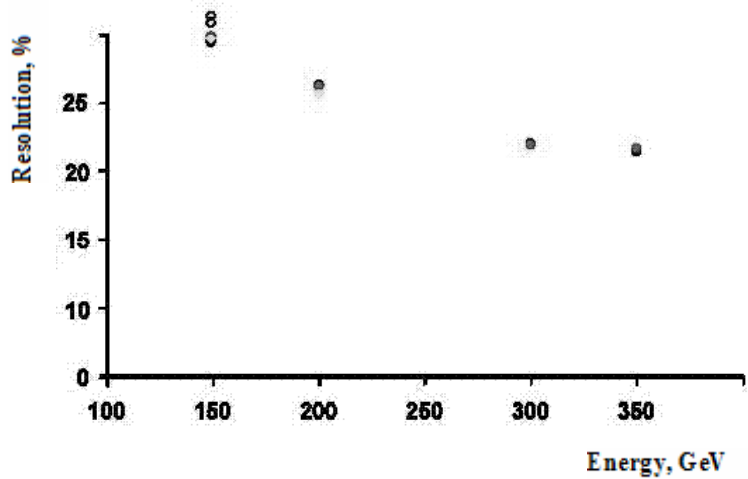

**Figure 13.** Positive pion energy resolution of the combined zero degree calorimeter.

## 8. Current status and summary
The installation of the ZDCs in the LHC sector 4-5 and sector 5-6 was finished in May 2008. A photograph of the ZDC installed in sector 4-5 is shown in figure 14. Both detectors were connected to read-out electronics and tested. Commissioning of detectors as a part of the CMS experiment and as a part of the LHC beam monitoring system has begun.

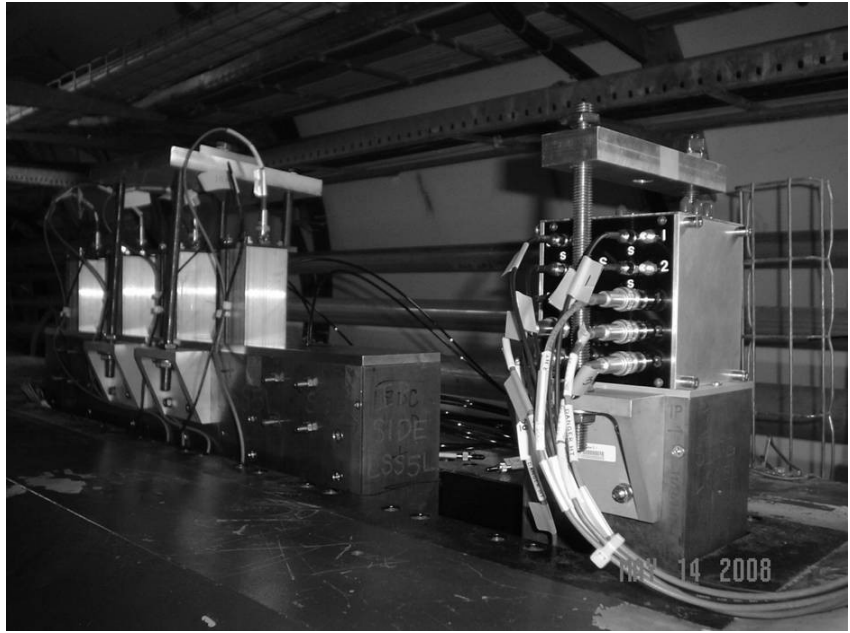

**Figure 14**. ZDC installed in sector 4-5 of the LHC.

**8.1 Summary**
- We have designed, constructed and tested the combined zero degree calorimeter;
- GEANT4 simulation reproduced well the test beam data;
- The calorimeter fulfils the technical, geometrical and physics requirements;
- The ZDC has sufficient energy resolution and linearity to meet our physics goals.   I

**Acknowledgments**
The detector studies were possible only because of the help of the CMS Collaboration and the SPS crew at CERN. It is our pleasure to acknowledge their important contribution.